\documentclass[epj,nopacs]{svjour} 
\pdfoutput=1
\usepackage{hyperref}
\usepackage{amsmath,slashed}
\usepackage{graphicx}
\usepackage{color}
\usepackage{hepunits}
\usepackage{amssymb}
\newcommand{\be}{\begin{eqnarray*}}
\newcommand{\ee}{\end{eqnarray*}}

\newcommand{\bee}{\begin{eqnarray}}
\newcommand{\eee}{\end{eqnarray}}
\newcommand{\beeq}{\begin{equation}}
\newcommand{\eeeq}{\end{equation}}

\DeclareGraphicsExtensions{.pdf,.png,.jpg}
\begin{document}
\title{Fractal based observables to probe jet substructure of quarks and gluons}
\author{Joe Davighi\inst{1} \and Philip Harris\inst{2}}
\institute{Department of Applied Mathematics and Theoretical Physics, University of Cambridge, Wilberforce Road, Cambridge, UK \and CERN, European Organization for Nuclear Research, Geneva, Switzerland}

\abstract{
New jet observables are defined which characterize both fractal
and scale-dependent contributions to the distribution of hadrons in
a jet. These infrared safe observables, named Extended Fractal Observables (EFOs), 
have been applied to quark-gluon
discrimination to demonstrate their potential utility. 
The EFOs are found to be individually discriminating
and only weakly correlated to variables used in existing discriminators.
Consequently, their inclusion improves discriminator performance,
as here demonstrated with particle level simulation from
the parton shower.
}

\maketitle

\section{Introduction\label{sec:Introduction}}

A hadronic jet is produced from an initial parton via a sequence of
perturbative QCD branching interactions (the parton shower),
followed by the non-perturbative conversion of partons to the hadrons we observe in experiments (hadronization).
A Markov chain description of the parton shower suggests the spatial distribution of partons will
exhibit some fractal character~\cite{Gustafson:1991ru,PhysRevD.45.4077,Andersson:1988ee,Larkoski:2012eh,Jankowiak:2012na,Soper:2011cr}, 
and this will be inherited by the final hadron distribution 
(invoking local parton-hadron duality ~\cite{0954-3899-17-10-017}). However, true scale invariance of the hadron distribution within a jet is
broken by the running of the branching probability, 
termination of the shower due to hadronization, and finite detector resolution.
Here we define new observables to characterize
jet branching structure, named Extended Fractal Observables (EFOs), which accommodate deviations from fractal structure through
simple parametrizations.
The idea is to apply box-counting techniques,
used widely in the study of dynamical systems and scale invariant
objects, to the substructure of QCD jets. Box counting has previously been employed in particle physics to
calculate the fractal dimension of electromagnetic showers~\cite{Ruan:2013iaa} 
for highly granular calorimetric reconstruction. Here, we
extend the generality and information content of this technique in our characterization of QCD jets.

The motivation for this study is two-fold. Firstly, we would like to
characterize the spatial substructure of jets into a set of new observables. Secondly, we would like to demonstrate the
use of such observables in the discrimination of quark and
gluon jets. Quark and gluon discrimination has long been used as a
tool to enhance the sensitivity of signatures with additional
quarks \cite{Aad:2014gea,CMS-PAS-JME-13-002,Badger:2016bpw,Gras:2017jty}. In particular, weak boson fusion induced Higgs-production is 
enhanced due to the distinct signature of two additional hard quark jets in the
gluon-dominated forward region of the detector \cite{FerreiradeLima:2016gcz,Gallicchio:2012ez,Gallicchio:2011xq,Gallicchio:2011xc,Abreu:1999rs,Briere:2007ch,Pumplin:1992bv,Seymour:1996np,Aad:2014gea,Kilic:2008ub}. Quark and gluon tagging are also expected to be useful for physics searches \textit{beyond} the Standard
Model, including the detection of supersymmetric particles~\cite{Bhattacherjee:2016bpy,Joshi:2012pu}. Additionally, if well designed, these taggers can be further extended to the subjets of boosted boson signatures~\cite{CMS-PAS-JME-14-002}. 
We demonstrate that modest improvements can be made to existing quark-gluon taggers by incorporating the new jet observables defined in this paper. 

Finally, our construction of pixel-based jet observables resonates with the recent development of the jet image paradigm ~\cite{Komiske:2016rsd,Cogan:2014oua}, 
in which the energy measured in each detector cell is interpreted as the intensity of a pixel in a 2D image. Within this approach, powerful machine-learning 
algorithms for classifying images have been brought to bear on a range of jet classification problems. This has included tagging
boosted weak bosons \cite{deOliveira:2015xxd,Cogan:2014oua}, boosted top quarks \cite{Almeida:2015jua}, and heavy-flavors \cite{Baldi:2016fql,Guest:2016iqz}. 

We define EFOs in the following section. In 
section \ref{sec:Performance-in-quark-gluon} we analyze the
performance of these observables in quark-gluon discrimination, before
concluding.

\section{Extended Fractal Observables\label{sec:Defining-BLF-and}}

The computation of the EFOs is performed on a jet by jet
basis using a variation of the Minkowski-Bouligand (box-counting) dimension, as follows. 

\begin{center}
\begin{figure*}[ht!]
\includegraphics[width=0.95\textwidth]{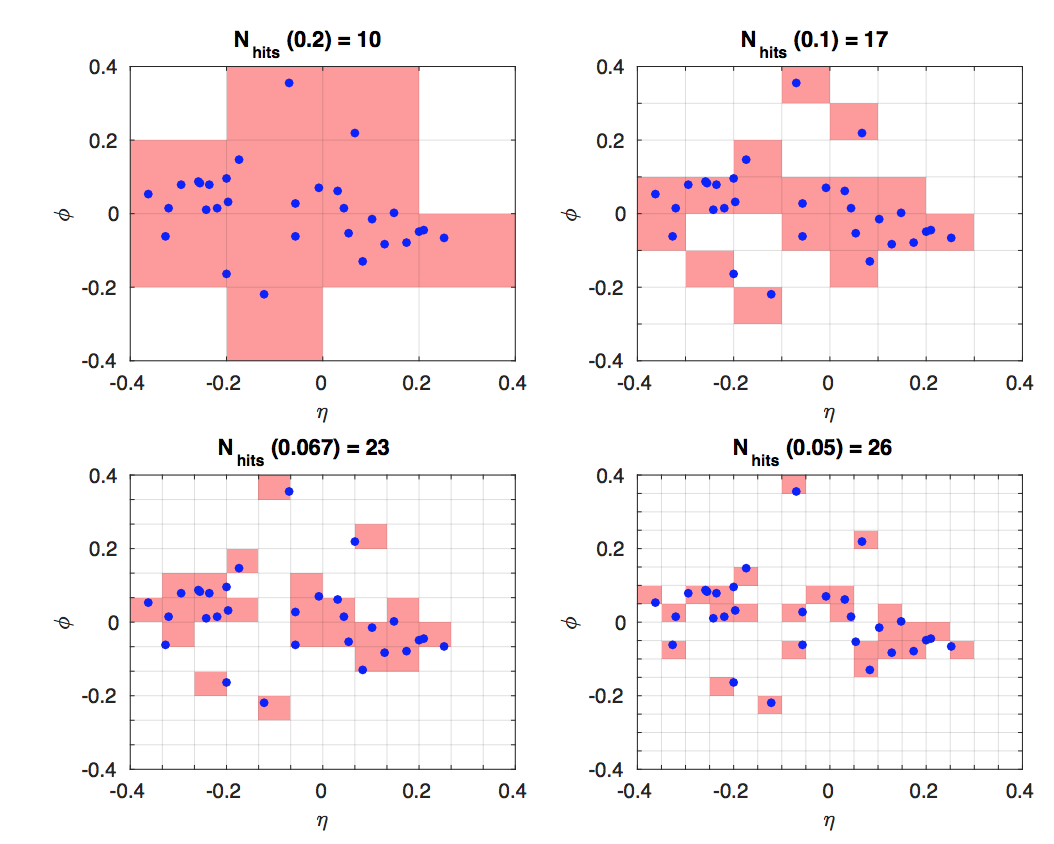}
\caption{
An illustration of the iterated box-counting procedure used to calculate fractal-based quantities on a set of points.
The filled blue circles are the $\left(\eta,\phi\right)$ angular coordinates of the hadrons within a particular
sample jet (in particular, this jet has total $p_{T}=157$~\GeV, and 30 constituent hadrons). The box-counting is 
illustrated for four sample scales, corresponding to successively finer $\epsilon$ values of $0.2$, $0.1$, $0.067$ and $0.05$. 
The cells registering particle hits are highlighted with red shading.
}
\label{fig:boxcounting}
\end{figure*}
\end{center}

\subsection{Variable definitions \label{subsec:variable defns}}

To define our variables we implement a two-stage recipe: firstly, the jet cone is divided
in the familiar $\left(\eta,\phi\right)$ angular coordinates into a square
grid of cells, each cell having side-length $\epsilon$. For a given
scale $\epsilon$, we count the number of cells $N_{hits}\left(\epsilon\right)$
which register particle hits with a total transverse momentum greater than some 
pixel-level soft cutoff, in this study chosen to be $p_{T}>1.0$~\GeV. This low energy cut
represents a limiting threshold due to detector resolution. This counting is iterated over a
range of scales, as is illustrated in Figure~\ref{fig:boxcounting}. The second stage is to fit smooth functions to
the variation of $y=\log N_{hits}\left(\epsilon\right)$ with $x=\log\left(1/\epsilon\right)$,
and to extract the parameters of the fit as a set of (correlated)
jet observables, which we call Extended Fractal Observables (EFOs). This is a generalization of the traditional box-counting
method, in which only linear functions $y=mx+c$ are fitted, with the gradient $m$ identified
as the fractal dimension~\cite{Ruan:2013iaa}.

\begin{center}
\begin{figure*}[t]
\includegraphics[width=0.45\textwidth]{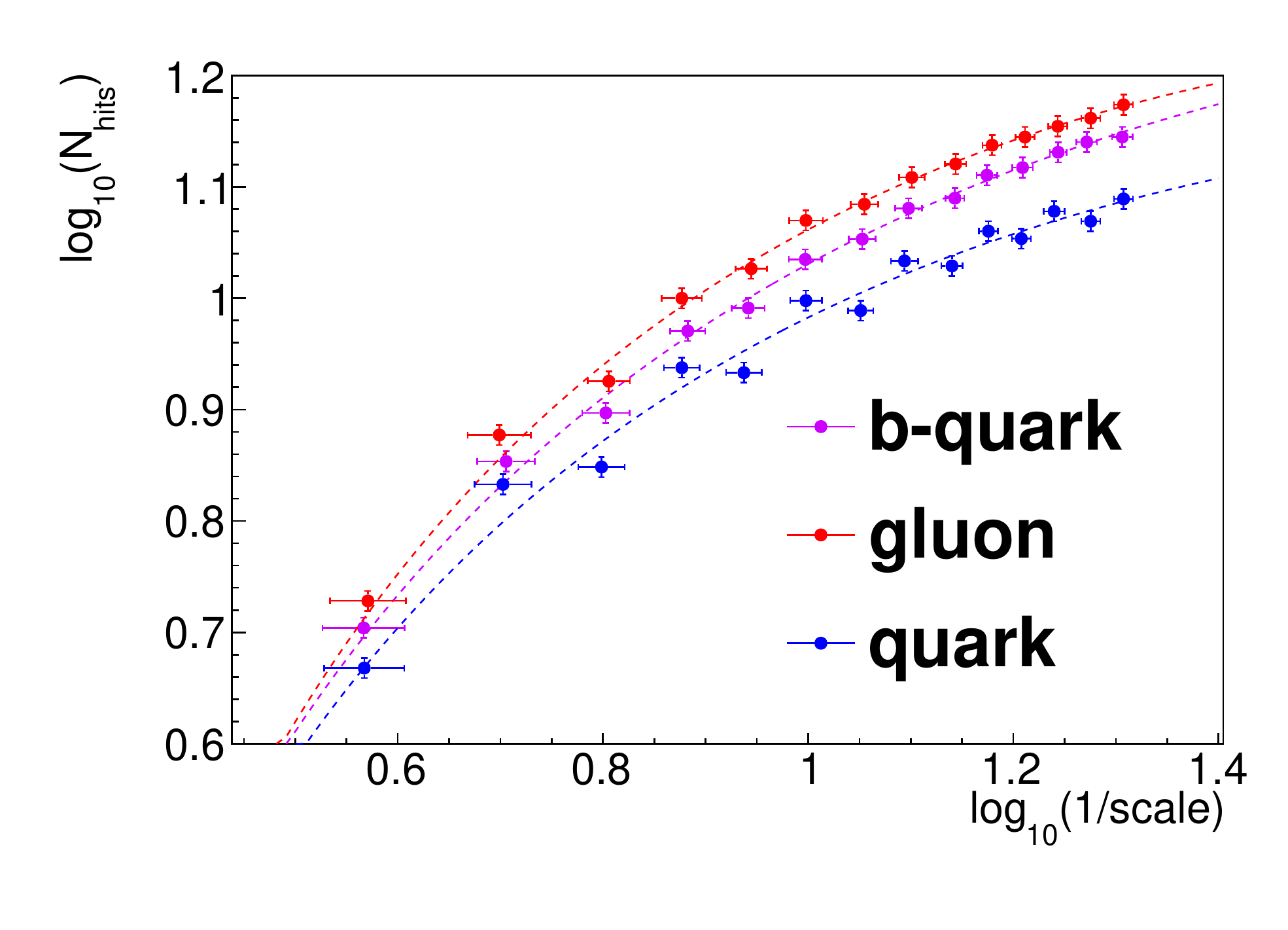}
\includegraphics[width=0.45\textwidth]{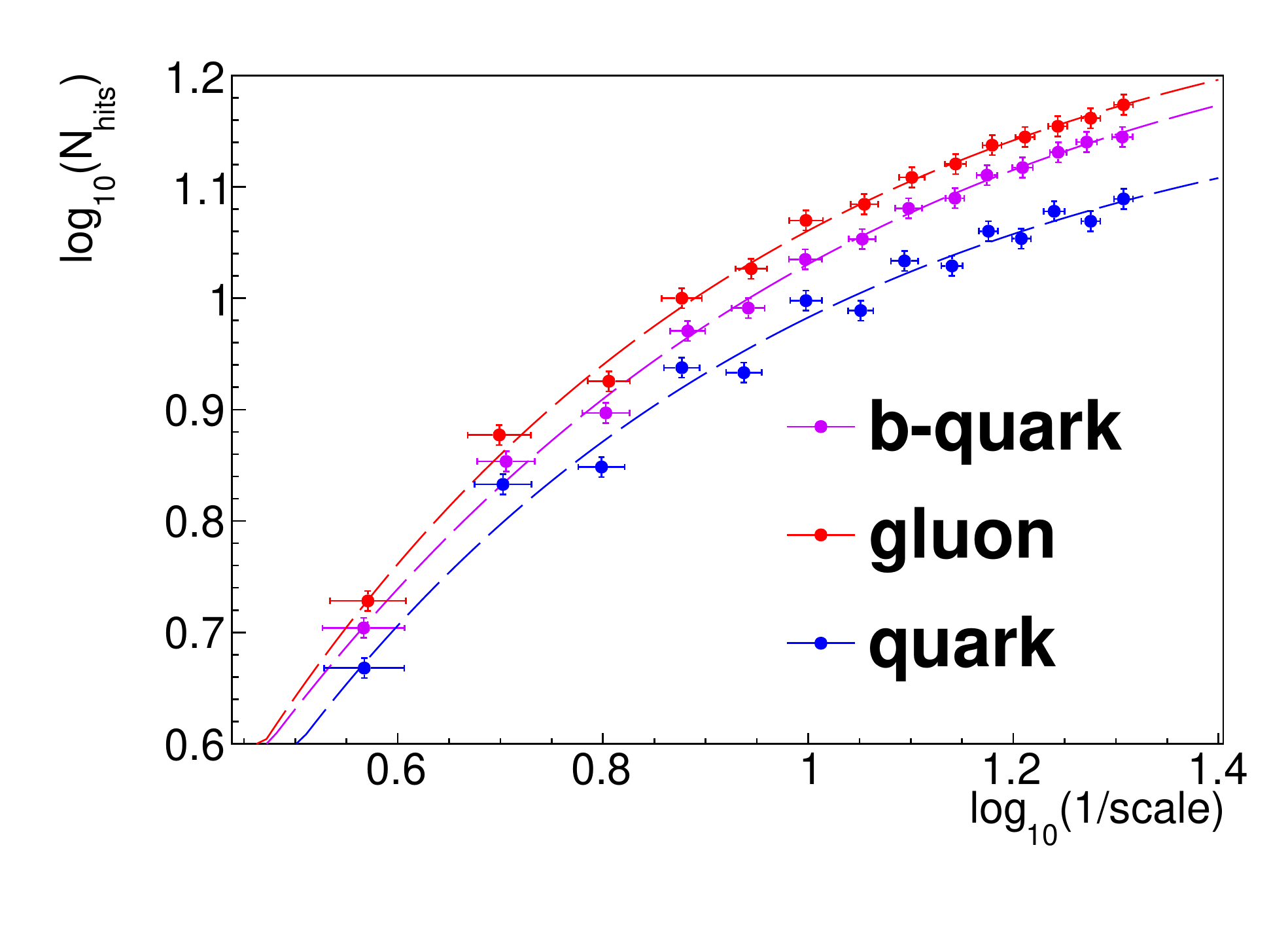}
\caption{Left: logarithmic fits to $\log N_{hits}\left(\epsilon\right)$ against $\log\left(1/\epsilon\right)$
for light quarks, bottom quarks, and gluons, of the form $y=p_{0}+p_{1}x+p_{2}\log x$. The values of the fitted parameters $\left\{ p_{i}\right\} $
define one possible set of Extended Fractal Observables.
Right: fits to $\log(N_{hits})$ against $\log\left(1/\epsilon\right)$ using an asymptotically saturating fitting function, specifically $y=p_{0}+p_{1}\tanh (x-p_{2})$.}
\label{fig:logquadfits}
\end{figure*}
\end{center}

\begin{center}
\begin{figure*}[t]
\includegraphics[width=0.45\textwidth]{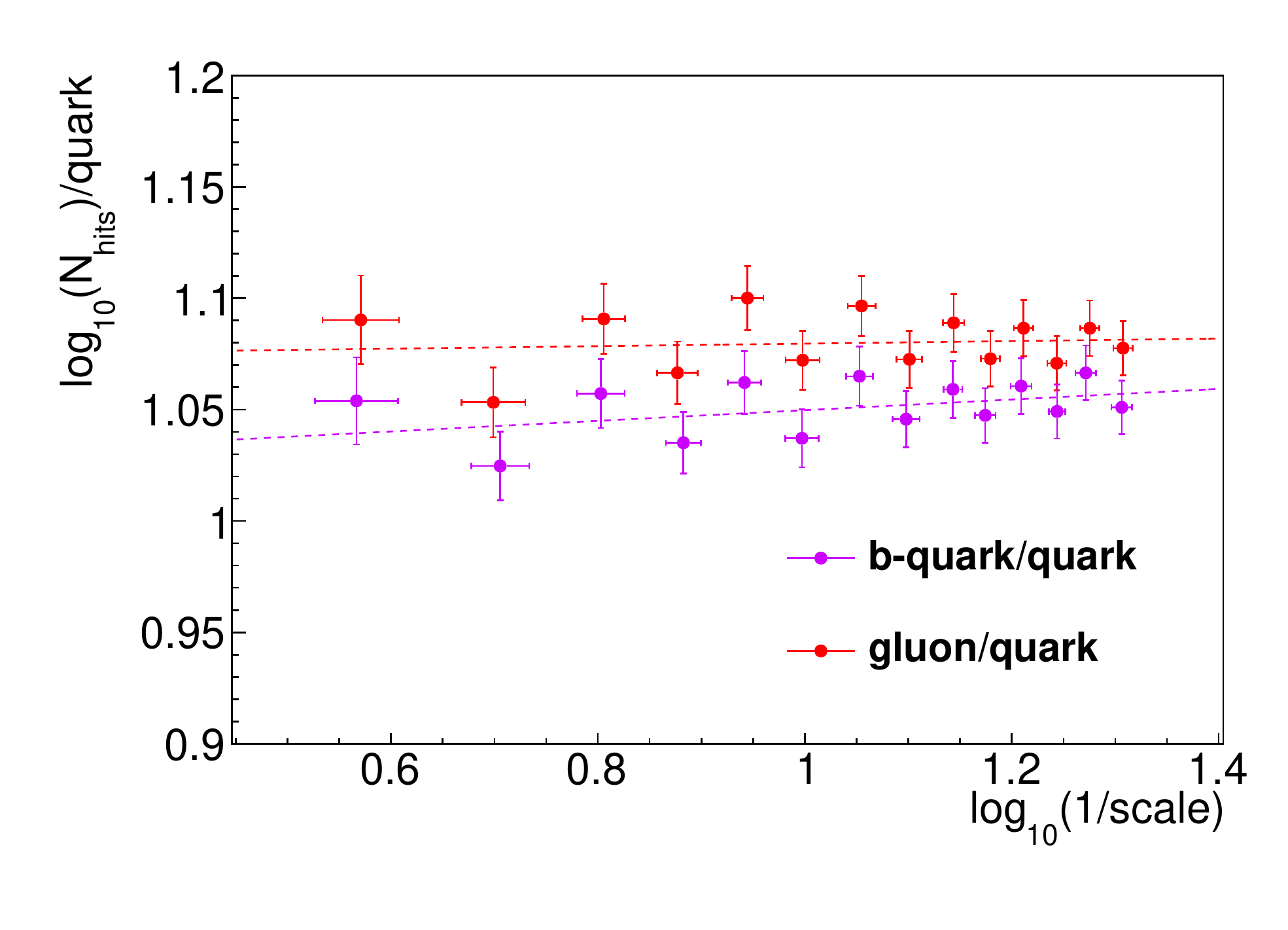}
\includegraphics[width=0.45\textwidth]{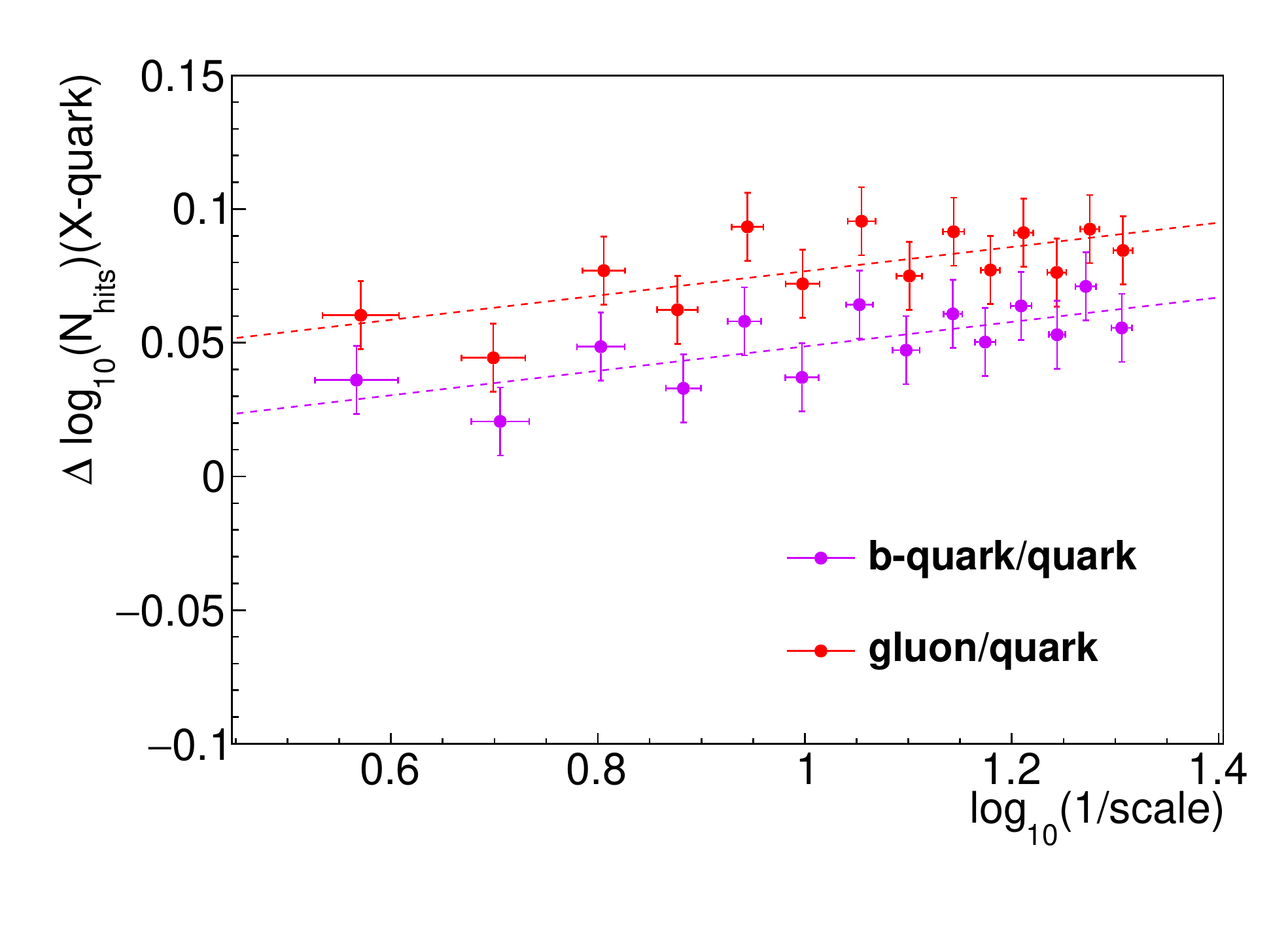}
\caption{Left: the \textit{ratio} of $\log(N_{hits})$ with respect to the quark values, for b-quarks and gluons, as a function of $\log\left(1/\epsilon\right)$. A linear fit is added for comparison.
Right: the \textit{difference} of $\log(N_{hits})$ with respect to the quark values, for b-quarks and gluons. In the Modified Leading Logarithmic Approximation (MLLA), the differences in hadron multiplicity between quarks, b-quarks and gluons are predicted to be energy independent ~\cite{Dokshitzer:2005ri}. The small but non-zero slopes in this plot reflect the fact that box-counting at a given angular scale probes spatial information in addition to the rate of splitting at the corresponding energy scale.}
\label{fig:comparisonfits}
\end{figure*}
\end{center}

Indeed, in Figure~\ref{fig:logquadfits} there is no distinct region of linear
scaling, as would be needed to extract a fractal dimension. Rather, $\log N_{hits}\left(\epsilon\right)$ levels off smoothly from large
to small scales as saturation is approached, motivating a non-linear fit to extract whatever information this curve might encode about the jet. In particular, the hadronization region (i.e. at small $\epsilon$) obviously carries non-perturbative information sensitive to the flavor of the jet. 
The observed curves are distinct between quarks, gluons and b-quarks, as summarized in Figures~\ref{fig:logquadfits} and ~\ref{fig:comparisonfits}. This scaling is a fundamental property of QCD resulting from the differences in the splitting of quarks and gluons. Further measurements of this scaling allows for an alternative approach to extract QCD properties such as the strong coupling constant~\cite{Bolzoni:2013rsa,Bolzoni:2012ii}.

The generic plateauing curves in Figure~\ref{fig:logquadfits} can be fitted by almost any non-linear function (given a suitably restricted range in $x$), so we studied fit functions with at most three parameters, for speed and robustness of fitting. Fits were carried out simply by a binned $\chi^2$ minimization of the chosen function. Example fit functions included the following:
\begin{enumerate}
\item logarithmic fits of the form $y=p_{0}+p_{1}x+p_{2}\log x$.
\item quadratic fits: $y=p_{0}+p_{1}x+p_{2}x^{2}$.
\item hyperbolic tangent fits: $y=p_{0}+p_{1}\tanh (x-p_{2})$.
\end{enumerate}
The values of the best fit parameters  $\left\{ p_{i}\right\}$ for each fitting function constitute three possible sets of EFOs. For a polynomial in $x=\log(1/\epsilon)$, like the quadratic fit function, the fit reduces to a matrix inversion and thus has a well-defined convergence. The other two parametrizations are not polynomials, hence we perform a $\chi^2$ minimization.

Functions which actually saturate, such as the hyperbolic tangent parametrization above, are more physically motivated because they can model the saturation itself (asymptoting to the jet multiplicity). However, for the range of box scales used in our study (of width $\epsilon \geq 0.05$, - see \ref{subsec:range of scales} below), and for all but the lowest $p_T$ jets, the \textit{non}-saturating fit functions also provide adequate models for the observed scaling. For the purpose of quark-gluon discrimination (see section ~\ref{sec:Performance-in-quark-gluon}), the logarithmic fitting function was found to give the best discrimination performance of the three functions above (see Figure \ref{fig:gain} to compare the performance between the logarithmic and hyperbolic tangent fitting functions).

\subsection{The range of box-counting scales \label{subsec:range of scales}}

The range of angular scales $\epsilon$ has been chosen by paving the jet
cone with a square grid of $N\times N$ cells, where the splitting scale $N$ ranges in
integer steps from 3 to 16. For each $N$, the angular scale is $\epsilon=2R/N$, where $R$ is the jet radius, in this study $R=0.4$.
The coarsest $\epsilon$ scale chosen, corresponding to $N=3$, 
is essentially the coarsest scale carrying potentially discriminating information (for $N=2$ 
the jet cone would be divided into four quarters, all of which will register a hit for realistic jet shapes). 
The finest $\epsilon$ scale chosen is
 $\epsilon_{min}=0.8/16=0.05$, because this is approximately the angular detector
resolution in both LHC experiments, CMS and
ATLAS~\cite{Aad:2008zzm,Chatrchyan:2008aa}. For the $p_{T}\geq 100$~\GeV~jets
studied here, the number of hits is just beginning to saturate at this scale (see Figure \ref{fig:logquadfits}), so we are probing into the hadronization region prior to the flat plateau. 

Finally, we would like to highlight that these fractal-based
observables are similar in spirit to calculating subjet rates of jets
~\cite{Gallicchio:2011xq,Bhattacherjee:2015psa}, given subjets clustered using
the $p_{T}$-independent Cambridge-Aachen algorithm \cite{Dokshitzer:1997in}. Both observables compute $p_{T}$-independent branching information on a
succession of angular scales down to some threshold. And both observables perform 
what is essentially a further clustering on the substructure of the jet to extract this information
pertaining to the branching history of the jet. 
In light of this, the EFO approach could be
extended to utilize subjet counts (instead of hit grid cell counts) to assign scale-dependent multiplicities $N(\epsilon)$. 

\subsection{Infrared and Collinear safety \label{subsec:IRC safety}}

Preserving infrared and collinear safety ensure calculability in
perturbative QCD. An observable is infrared (collinear) safe if its value is
unchanged by the emission of soft (co-moving) particles. The EFOs, as defined in
\ref{subsec:variable defns} with a pixel-level soft cutoff, are fully IRC safe.

Firstly, the box counting procedure is intrinsically collinear safe: if one particle
splits into two particles with the same $\left(\eta,\phi\right)$
coordinates, we still count just one cell hit by both daughter
particles, at any finite scale of probing. Hence
collinear splittings will not affect the number of cells $N_{hits}\left(\epsilon\right)$
to register particle hits at any choice of scale. On the other hand, infrared safety of the EFOs can only be engineered by
imposing some low momentum cutoff to cleanse the jet of its soft constituents.  However, this soft cutoff must be implemented consistently with collinear safety.
If we simply discarded all soft hadrons with, say $p_{T}<1$~\GeV, this would spoil collinear safety. 
To see this, consider the following pathological example: if a particle with $p_{T}=1.5$~\GeV~ 
splits into two comoving particles with $p_{T}=0.8$~\GeV~ and $p_{T}=0.7$~\GeV, then
both would be discarded by a particle-level soft cut, and so $N_{hits}\left(\epsilon\right)$
would not be invariant under this collinear splitting.

This is remedied by defining a pixel-level (rather than particle-level) sort cutoff. That is, we only consider a cell
to register a hit if it measures a total $p_{T}$ greater than our soft cutoff of $1$~\GeV. This way, if the troublesome
$1.5$~\GeV particle in the example above splits collinearly into any number of daughters, the pixel still measures a total $p_{T}$ of $1.5$~\GeV, and 
so registers a hit regardless of these splittings. Thus, box-counting with a pixel-level soft cutoff is fully IRC safe. In addition, a pixel-level rather than particle-level cut is more 
naturally realized experimentally since a pixel hit is consistent with
an LHC detector cell. 

Numerically, the performance of a quark-gluon discriminant built using the EFOs was found to be essentially insensitive to varying the value of this $p_{T}$ cut (over values between $0.1$~\GeV \ and $1.0$~\GeV), suggesting the variables are not strongly shaped by the IR emission, at least in simulations. In the following section, a $p_{T}$ cut of $1$~\GeV \ is used throughout. Finally, we acknowledge that pixel-level cutoffs have been used previously in the context of jet images analyses (for example in ~\cite{Komiske:2016rsd}) to ensure IRC safety in the same context.

\section{Performance in Quark-Gluon Discrimination\label{sec:Performance-in-quark-gluon}}

We now investigate whether these observables might be a useful new tool in the important and challenging problem of distinguishing light quarks from gluon jets.

\subsection{Event generation and setup} \label{subsec:setup}

In this study, we use QCD dijet samples at a center-of-mass energy of 13~\TeV. Because previous quark-gluon studies have revealed that discrimination performance varies a lot between the different generators ~\cite{Larkoski:2014pca,Aad:2014gea,CMS-PAS-JME-13-002,Gallicchio:2012ez,Badger:2016bpw}\footnote{Herwig has been consistently seen to give the more conservative estimates of discrimination power, both with respect to Pythia and real LHC data.}, we here produce and shower events (at leading order) using both Herwig++ (version 2.7.0 with tune UE-EE-5C ) ~\cite{Bahr:2008pv,Seymour:2013qka} and Pythia 8 (version 8.185 with tune CUETP8M1)\cite{Khachatryan:2015pea},  with order 150k events in each.
Jets are clustered with the anti-$k_{T}$ algorithm using the final state particles following showering and hadronization; a cone size of $R=0.4$ and the FastJet code package~\cite{Cacciari:2011ma}  are used for the jet clustering. The EFOs (here computed using the logarithmic fitting function), along with a set of other established jet observables, have been computed for the highest $p_{T}$ jet in each event. We define the flavor of that jet by matching to the highest-$p_{T}$ parton within $R<0.3$ of the jet axis, and classify the event as signal (background) if matched to a light quark (gluon)\footnote{Note that b(bottom)-jets may be efficiently identified using a secondary vertex tagger, and separately vetoed.}.

\begin{figure*}[t]
\begin{center}
\includegraphics[width=0.46\textwidth]{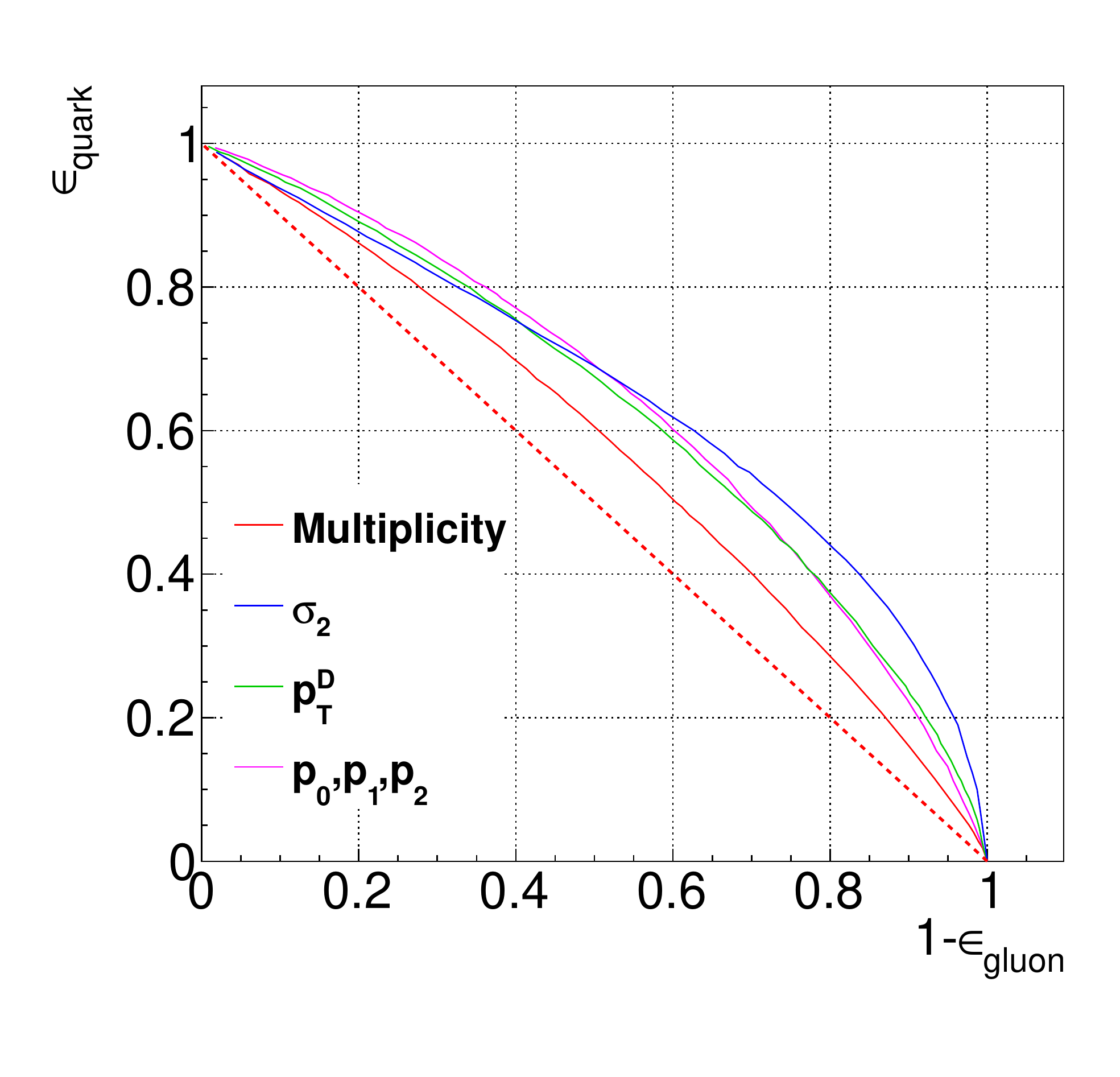}
\includegraphics[width=0.46\textwidth]{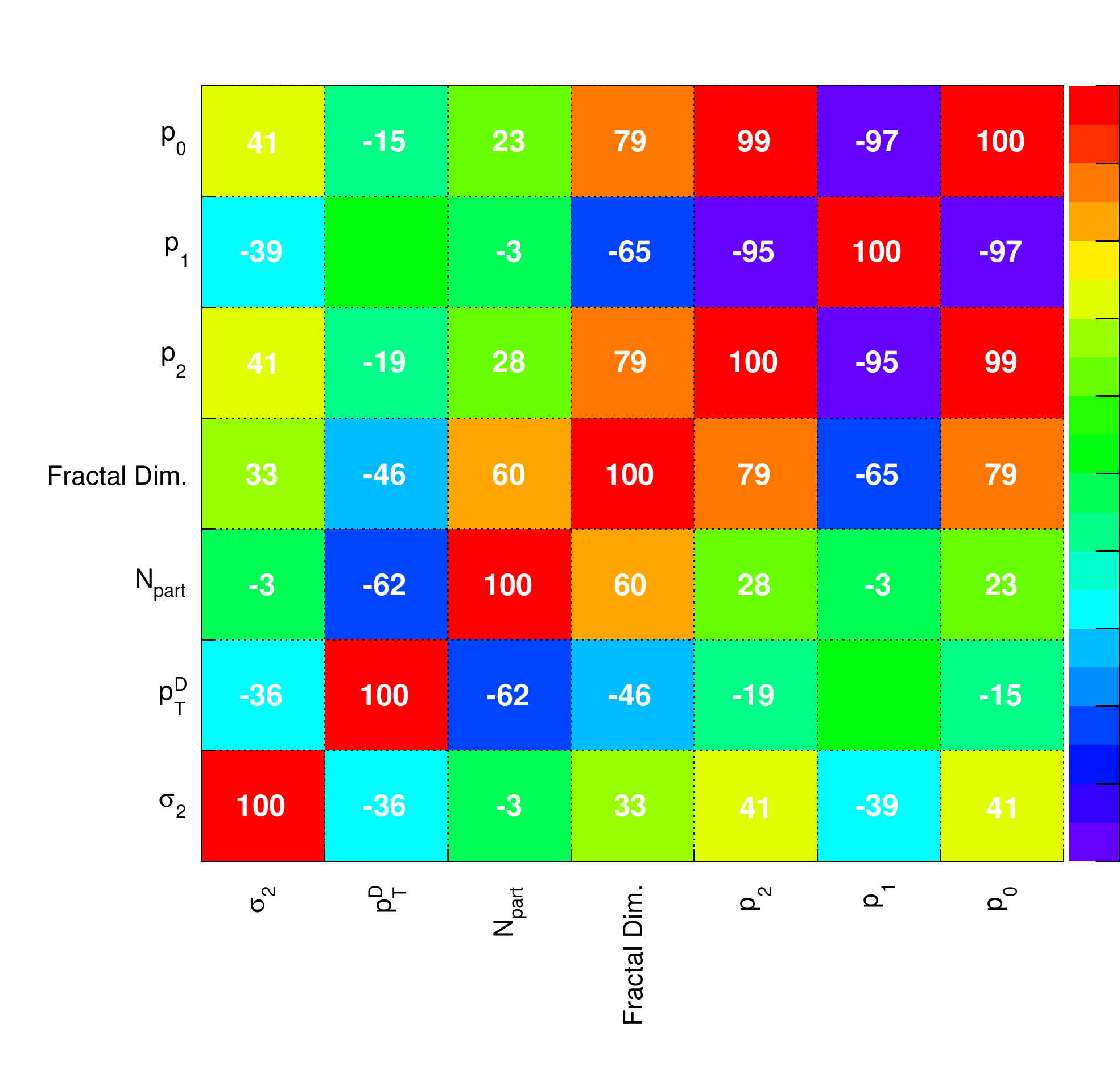} \\
\caption{Left: single variable performance ROC curves. The EFOs, minor axis, and $p_{T}D$ are
significantly more discriminating than multiplicity. The EFOs
are most discriminating for high signal efficiency ($\gtrsim70\%$),
below which jet minor axis becomes most discriminating.
Right: linear correlation coefficients
between pairs of variables, for quark jets (the values are similar for gluon jets). We see only weak correlations between the EFOs and the three
existing QGD variables.}
\label{fig:singlevariables}
\end{center}
\end{figure*}

As a baseline for comparison, we shall consider the variables currently used by the Compact Muon Solenoid (CMS) quark-gluon
tagger, which are \cite{CMS-PAS-JME-13-002}: i) the total number of reconstructed particles in the jet
(the multiplicity) \cite{Alexander:1995bk}; ii) the $p_{T}D$ variable ($C_{1}^{\beta=0}$)\cite{Larkoski:2013eya}, 
\begin{equation}
p_{T}D=\frac{\sqrt{\Sigma_{i}p_{T,i}^{2}}}{\Sigma_{i}p_{T,i}},
\end{equation}
where $i$ sums over the constituents of the jet, which describes the distribution of transverse momentum between the particles in the jet; and
iii) $\sigma_{2}$, the ($p_{T}$-weighted) semi-minor axis of the jet in the $(\eta,\phi)$ plane \cite{CMS-PAS-JME-13-002}, defined by
\begin{equation}
\sigma_{2}=(\lambda_{2} / \Sigma_{i}p_{T,i}^{2})^{1/2},
\end{equation}
where $\lambda_{2}$ is the smaller eigenvalue of the $2\times 2$ symmetric matrix with components
$M_{11}=\Sigma_{i} p_{T,i}^{2} \Delta \eta_{i}^{2}$, 
$M_{22}=\Sigma_{i} p_{T,i}^{2} \Delta \phi_{i}^{2}$, and
$M_{12}=-\Sigma_{i} p_{T,i}^{2} \Delta \eta_{i} \Delta \phi_{i}$.  
Throughout this study, we build multi-variable quark-gluon discriminants using a boosted decision tree (BDT), implemented using the Toolkit for Multivariate Analysis (TMVA) via adaptive boosting. The $p_{T}$ of the quark and gluon samples are reweighted to match the exact same kinematics in both cases, so as to avoid selection biases induced by  kinematic differences in the simulation.

\subsection{Results}

\begin{figure*}[ht]
\begin{center}
\includegraphics[width=0.75\textwidth]{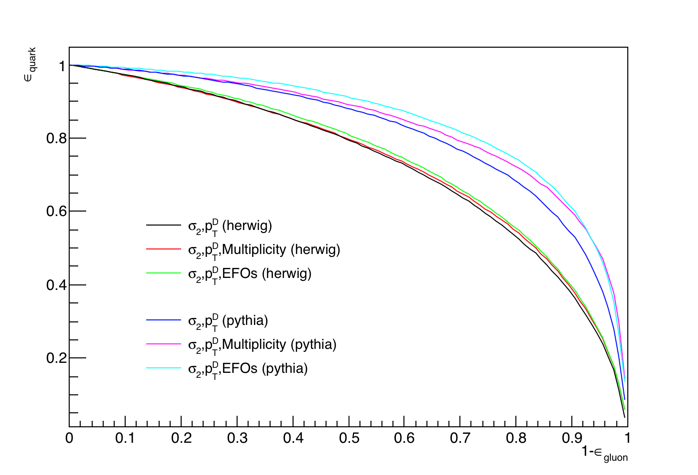}
\caption{ROC curves for BDT discriminators constructed from various combinations of observables, as indicated by the legend, for events showered using both Herwig and Pythia with jet $p_{T}\geq 100$~\GeV. The discrimination is superior in Pythia. We see in both event generators that including the EFOs rather than multiplicity (which is used in the CMS tagger) yields a marginally better performance. 
}
\label{fig:newtagger}
\end{center}
\end{figure*}

\begin{figure*}[ht]
\begin{center}
\includegraphics[width=0.49\textwidth]{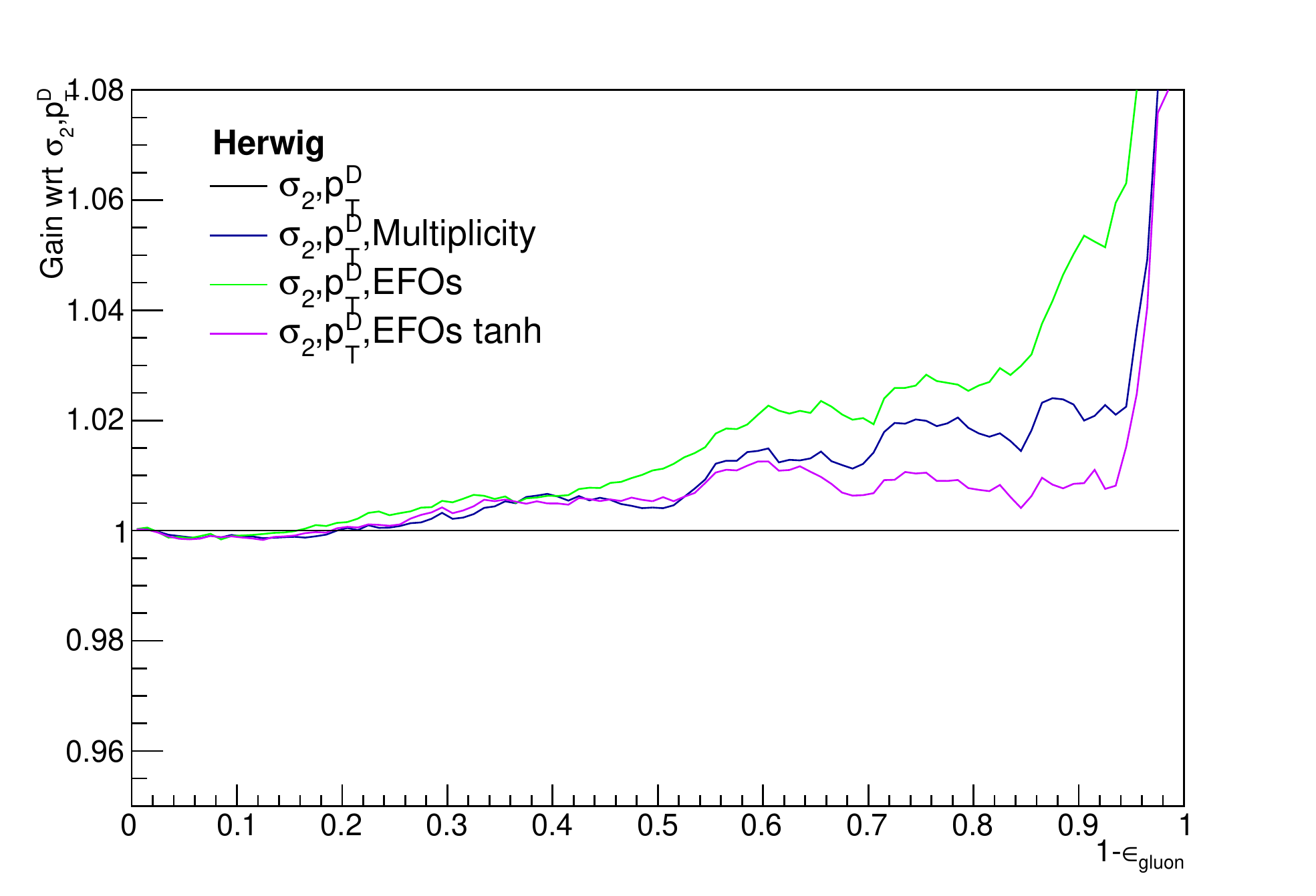}
\includegraphics[width=0.49\textwidth]{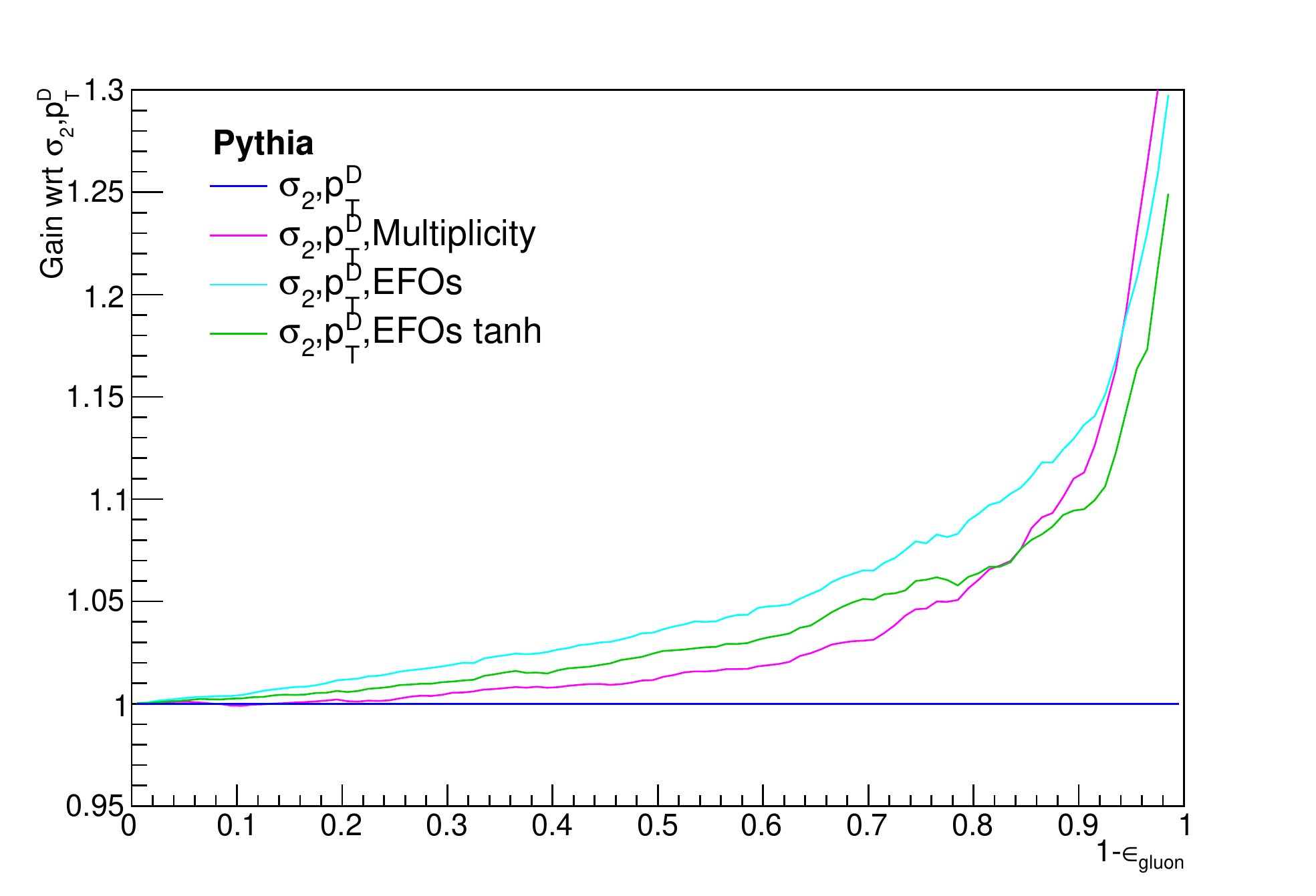}
\caption{Left: the relative gain for the three-variable taggers with respect to a baseline tagger using just $p_{T}D$ and $\sigma_2$, for the Herwig events (which yield more conservative discrimination estimates). The gain is also plotted for EFOs computed with the hyperbolic tangent fitting function specified in subsection \ref{subsec:variable defns}, for which the performance is worse.
Right: for Pythia events. Note the wider range of the y-axis, to accommodate the larger gains found in Pythia.}
\label{fig:gain}
\end{center}
\end{figure*}

We first compare the discriminator performance of single variables and the correlations between them, before going on to compare multi-variable taggers built with and without inclusion of the new EFO observables.

We can measure discriminator performance by receiver operator characteristic (ROC) curves, which plot background rejection against signal efficiency. Roughly speaking, the more convex the curve, the better the performance. The left plot of Figure \ref{fig:singlevariables}, made using the Herwig samples, shows that the EFOs \footnote{We use a BDT discriminator built from the combination of the three EFOs, $p_0$, $p_1$ and $p_2$. While the combination of all three EFOs adds little discrimination beyond that of a single EFO due to their near-perfect correlation,
the selection of any single $p_i$ would be arbitrary for the sake of this comparison.} are individually well-discriminating, particularly if we seek high signal efficiency. Their performance is significantly better than that of the jet multiplicity variable. 

The right plot of Figure \ref{fig:singlevariables} presents the linear correlation coefficients (calculated using the TMVA toolkit) between the EFOs and the existing CMS quark-gluon tagger variables: multiplicity, $p_{T}D$ and $\sigma_{2}$. We also include a computation of the fractal dimension, which has been calculated from a linear fit over a small range of box scales. Strong correlations are present amongst the EFOs, as is natural given they are parameters derived from the same fit. However, their correlations with the other variables are no greater than 43\% (for either quarks or gluons)\footnote{Note that the traditional fractal dimension is more strongly correlated to existing QGD variables, particularly multiplicity.}. Interestingly, the EFOs are most highly correlated with $\sigma_{2}$, not multiplicity as might have been expected. This evidence suggests the discrimination power of the EFOs is not simply a result of higher multiplicities in gluon jets, and therefore that the addition of these parameters to a quark-gluon discriminator might improve performance.

We find that replacing the multiplicity variable in the existing CMS quark-gluon tagger with the EFO variable yields a gain in discriminator performance, albeit only a modest one. This gain is seen using both Herwig and Pythia event generators (with the setup described above) in the ROCs presented in Figure \ref{fig:newtagger}, which are for jets with $p_{T}\geq 100$~\GeV. We see the performance in Pythia is significantly better than Herwig for each combination of variables, consistent with previous studies ~\cite{Aad:2014gea,CMS-PAS-JME-13-002,Gallicchio:2012ez,Badger:2016bpw}. 

\begin{figure}[ht]
\begin{center}
\includegraphics[width=0.49\textwidth]{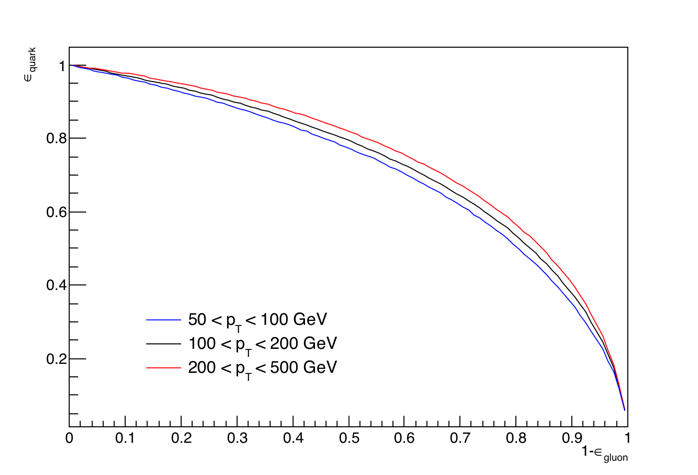}
\caption{Performance of a possible new quark-gluon tagger (using $p_{T}D$, $\sigma_2$, and the EFOs), in three $p_T$ bins, for Herwig-generated dijet events. Quarks and gluons are found to be easier to distinguish using this tagger at higher $p_T$.
}
\label{fig:ptbin}
\end{center}
\end{figure}

Moreover, the incremental gain upon replacing multiplicity with the EFOs is larger in Pythia than Herwig, so Herwig gives the more conservative estimate of the impact of including the EFOs. We see the gain in performance (relative to a baseline tagger using just $p_{T}D$ and $\sigma_2$) more clearly in Figure \ref{fig:gain}, with the left panel for Herwig and the right for Pythia. The gain is at the level of $1-2\%$ in the more conservative Herwig setup, and slightly larger in Pythia (note the different scaling of the y-axis). To emphasize a previous point, these gains were found to be stable across different values of the soft $p_{T}$ cut. Finally, we investigated how the performance varies with energy scale, by performing the analysis in $p_T$ bins of $50-100$~\GeV, $100-200$~\GeV, and $200-500$~\GeV. Discrimination was found to increase with $p_T$ in both Herwig and Pythia (see Figure \ref{fig:ptbin} for the Herwig results).

Combining all four variables (multiplicity, $p_{T}D$, $\sigma_2$ and the EFOs) was seen to give no further improvement. This suggests all the information from multiplicity is captured by the 
EFOs\footnote{This is unsurprising, because jet multiplicity is simply the asymptotic number of hits as we approach the saturation region.}, while the converse is not true. In summary, we have presented evidence in this study that the Extended Fractal Observables provide an additional handle that captures the salient features of jet multiplicity, incorporates new information from showering and hadronization, and which is also better behaved under IRC emission (see \ref{subsec:IRC safety}).

\section{Conclusions\label{sec:Discussion}}

In this study we defined new jet observables, the Extended Fractal Observables,
by a generalization of the box-counting method used in the study of
fractal systems. Defined with a pixel-level low momentum cutoff, 
these observables are infrared and collinear safe. We have then sought
to apply the EFOs to improve quark-gluon discrimination.
At the generator level, we find some modest improvement in discrimination
by gluon rejection when we replace multiplicity with the EFOs in the existing
CMS tagger, across both Herwig++ and Pythia 8. Extending the performance of these new variables to include
detector effects can naturally be performed in the LHC environment with the CMS Particle Flow algorithm
\cite{CMS-PAS-PFT-09-001} in conjunction with the PUPPI algorithm \cite{Bertolini:2014bba} to reconstruct
particle candidates in the presence of high pile-up. 

\section{Outlook\label{sec:outlook}}

This method of studying jet substructure is a new approach.
As such, there are many directions in which we would like to proceed, including:
\begin{enumerate}
\item Exploring particle hits in a 3-dimensional coordinate space spanned by $\eta$,
$\phi$ and $z^{-1}$, where $z$ is the fractional transverse momentum
of the jet constituent.
\item Applying the EFOs beyond  Quark-Gluon discrimination, for example to the identification of pile-up jets, or initial state radiation.
\item These box-counting methods extend very naturally from the substructure of a single
jet to a whole-event analysis. Such a novel approach may provide new
insight into searches for new physics topologies such as those in
supersymmetry or top quark pair production~\cite{Soper:2014rya}.
\item Furthermore, box-counting analyses could provide a useful characterization of event shapes in heavy ion collisions, 
where studies of jet properties beyond jet reconstruction are
traditionally difficult, but well motivated~\cite{Chatrchyan:2013kwa,CMS-PAS-HIN-15-004,CMS-PAS-HIN-16-006}.
\item Finally, we would like to emphasize that the calculation of EFOs on quark and gluon jets probes parton shower scaling that results from the QCD color factor ratio. Calculating EFOs on cosmic ray air shower profiles \cite{Brooijmans:2016lfv} could therefore help discriminate QCD-induced air showers from more interesting signals; of particular interest, showers induced by electroweak sphalerons. Experimentally, the calculation of EFOs in this air shower context is conceptually appealing: the 1660 individual Cerenkov detectors (spread over 3000 ~km$^2$) of the Pierre Auger Observatory in Argentina ~\cite{ThePierreAuger:2015rma} would naturally function as the finest-scale cells in our box-counting algorithm. These techniques could therefore be useful in probing physics at energies far beyond that of the LHC.
\end{enumerate} 

\section{Acknowledgments}
JD's work has been supported by The Cambridge Trust, and by the STFC consolidated grant ST/L000385/1.
We thank the CERN summer student program where
this work was initiated. We also thank Andrew Larkoski
for his insightful comments when performing these studies, and Bryan
Webber for helpful discussions. Finally, we thank Eric Metodiev for
helpful comments. 

\bibliography{articles_bib}
\bibliographystyle{JHEP}
\end{document}